\documentclass{emulateapj070207}
\usepackage{apjfonts}

\newcommand{\chandra}{{\em Chandra}}
\newcommand{\nancay}{{Nan\c{c}ay}}
\newcommand{\xte}{{XTE~J1810--197}}

\journalinfo{The Astrophysical Journal, in press}
\slugcomment{Received 2006 October 22; accepted 2007 March 19}

\shorttitle{XTE~J1810--197: VARYING TORQUE, RADIO FLUX AND PROFILES}
\shortauthors{CAMILO ET AL.}

\begin{document}

\title{The magnetar XTE~J1810--197: variations in torque, radio flux
density and pulse profile morphology}

\author{F.~Camilo,\altaffilmark{1} 
  I.~Cognard,\altaffilmark{2}
  S.~M.~Ransom,\altaffilmark{3} 
  J.~P.~Halpern,\altaffilmark{1}
  J.~Reynolds,\altaffilmark{4} 
  N.~Zimmerman,\altaffilmark{1} 
  E.~V.~Gotthelf,\altaffilmark{1}
  D.~J.~Helfand,\altaffilmark{1} 
  P.~Demorest,\altaffilmark{5} 
  G.~Theureau,\altaffilmark{2}
  and D.~C.~Backer\altaffilmark{5} }

\altaffiltext{1}{Columbia Astrophysics Laboratory, Columbia University,
  New York, NY 10027.}
\altaffiltext{2}{Laboratoire de Physique et Chimie de l'Environnement,
  CNRS, F-45071 Orleans, Cedex 2, France.}
\altaffiltext{3}{National Radio Astronomy Observatory, Charlottesville, 
  VA 22903.}
\altaffiltext{4}{Australia Telescope National Facility, CSIRO, Parkes
  Observatory, Parkes, NSW 2870, Australia.}
\altaffiltext{5}{Department of Astronomy, University of California,
  Berkeley, CA 94720-3411.}

\begin{abstract}
We report on 9 months of observations of the radio-emitting anomalous
X-ray pulsar \xte\ starting in 2006 May using the \nancay, Parkes,
GBT, and VLA telescopes mainly at a frequency of 1.4\,GHz.  The torque
experienced by the neutron star during this period, as inferred from a
measurement of its rotational frequency derivative, decreased by 60\%,
although not in a steady manner.  We have also observed very large ongoing
fluctuations in flux density and pulse shape.  Superimposed on these, a
general diminution of flux density and a broadening of the pulse profile
components occurred nearly contemporaneously with a decrease in torque of
about 10\% that took place in late 2006 July over an interval of 2 weeks.
After a slight increase in average flux density, since 2006 October the
flux density has continued to decline and the pulse profiles, while still
varying, appear more uniform.  In addition, a simultaneous observation of
the pulsar with the {\em Chandra X-ray Observatory\/} and the GBT allows
us to show how the X-ray and radio profiles are aligned.  We discuss
briefly the implications of these results for the magnetospheric currents
in this remarkable object.

\end{abstract}


\section{Introduction}\label{sec:intro}

The transient anomalous X-ray pulsar (AXP) \xte\ ($P=5.54$\,s) was
discovered in early 2003 when its X-ray luminosity increased $\sim
100$-fold \citep{ims+04} compared to the quiescent state maintained for
$> 24$ years \citep{hg05}.  Initial X-ray observations revealed unsteady
spin-down with $\dot P \approx 10^{-11}$ (implying a surface magnetic
dipole field strength $B \approx 2\times 10^{14}$\,G) that varied by up
to a factor of 2 within a few months of the outburst.  Such rotational
instability has been observed in other magnetars \citep[see][for a review
of the properties of magnetars]{wt06}.

The origin of a radio source positionally coincident with the pulsar
\citep{hgb+05} was clarified by the detection of strong, narrow, highly
linearly polarized radio pulses once per stellar rotation \citep{crh+06}.
This emission has several unique characteristics.  First, it is at
least one order of magnitude brighter than in the late 1990s, when
it was undetected; presumably the radio source turned on following
the X-ray outburst.  Just as the X-ray flux has now decayed to a level
approximating the historical low state \citep{gh06}, the radio emission
can be expected to cease ``shortly''.  In the initial observations,
\citet{crh+06} also noted that the radio flux density of \xte\ varied
on time scales of approximately a day in a manner inconsistent with
interstellar scintillation, implying large day-to-day radio luminosity
changes not seen in ordinary pulsars.  The ``average'' pulse profiles were
also found to change in a manner seemingly inconsistent with variations
observed in other neutron stars.  Lastly, the magnetar has a very flat
radio spectrum, and has been detected at higher radio frequencies than
any other pulsar.

These unusual characteristics of \xte\ presumably reflect
different physical conditions in its magnetosphere compared both
with those of ordinary radio pulsars and of persistent magnetars,
which so far have shown no evidence of magnetospheric radio activity
\citep[e.g.,][]{bri+06}.  The study of this radio emission thus provides
a new window into the coronae of magnetars.  We summarize here some
results from our monitoring program of \xte, focusing on rotational,
flux density, and pulse profile evolution.

\section{Data Acquisition, Analysis, and Results}\label{sec:data}

\subsection{Observations}\label{sec:obs}

\subsubsection{\nancay}\label{sec:nancay}

\begin{figure}
\begin{center}
\includegraphics[scale=0.50]{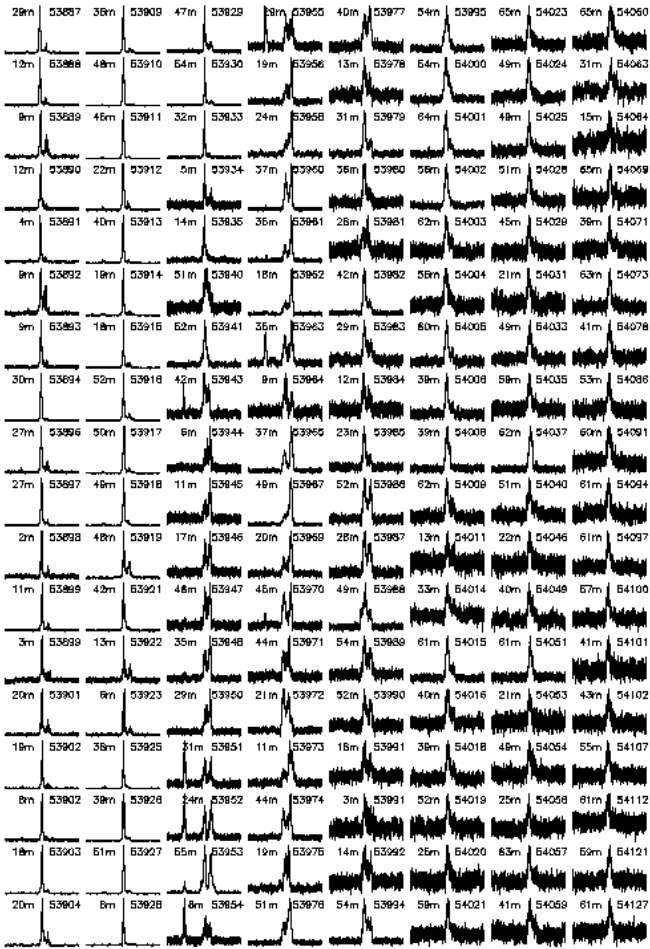}
\caption{\label{fig:profs_nancay} 
Daily pulse profiles from \xte\ recorded in the linear horizontal
polarization at \nancay\ at a frequency of 1.4\,GHz across a bandwidth
of 64\,MHz (\S~\ref{sec:nancay}), each displayed with 2048 bins (the
first is from 2006 June 1 and the last from 2007 January 27).  We aligned
these profiles using an ephemeris containing 11 frequency derivatives that
resulted in featureless residuals (see \S~\ref{sec:timing}).  Each profile
is labeled with the effective integration time (i.e., after the removal of
RFI) in minutes to its left and the observing date (MJD) to its right. In
some cases minor artifacts remain, due to incompletely excised RFI.
{\em A high-resolution version of this figure can be obtained at
http://www.astro.columbia.edu/$\sim$fernando/XTEJ1810-197Nancay.eps.}
}
\end{center}
\end{figure}

We have observed \xte\ at \nancay\ since 2006 June 1
(MJD~53887) with typical integration times of 15--60\,min,
using the Berkeley-Orleans-\nancay\ (BON) coherent dedispersor
\citep{ct06} based on a Serendip~V spectrometer\footnote{See
http://seti.berkeley.edu/casper/projects/SERENDIP5/}.  The \nancay\ radio
telescope has a gain of 1.5\,K\,Jy$^{-1}$ and a system temperature of
47\,K at 1.4\,GHz in the direction of \xte, which this meridian-type
telescope can track for 1\,hr each day.  We monitored the pulsar
nearly every day until 2006 October, and have done so every 2--3 days
on average since then.  The last observations reported on here are from
2007 January 27.  Dedispersion of a 64\,MHz band centered on 1398\,MHz
is done coherently into sixteen 4\,MHz channels using a 64-node computer
cluster, with each data stream then folded every 2\,min at the pulsar's
predicted period.  In Figure~\ref{fig:profs_nancay} we show the daily
folded pulse profiles obtained in the linear horizontal polarization (with
dipole orientation parallel to the ground) after excising significant
levels of radio frequency interference (RFI).  Regional telephone relay
and radio transmitters use the vertical polarization, leading to much
more severe RFI in that channel.  However, the radio emission from \xte\
is highly polarized \citep{crj+07} and fortuitously is much stronger in
the direction parallel to the ground at \nancay, so that the vertical
signal is in any case usually very small (the parallactic angle change
during the maximum integration is a negligible $\pm2\fdg5$).

Since 2006 July we have used a second observing system in parallel with
BON, where square-law detectors sampled every 2\,ms record the total power
for each polarization channel in a 50\,MHz band centered on 1410\,MHz.
These data are flux-calibrated by the firing of a calibrated pulsed noise
diode at the beginning of each observation.  However, this detector's
sampling rate is not stable enough for pulsar timing purposes, for which
we use the accurately time-tagged BON records.

We have used data from both systems to obtain flux density estimates for
the daily profiles.  The square-law detected outputs suffer less from RFI
(because of their higher central frequency and narrower bandwidth) and are
accurately flux-calibrated.  On the other hand, removal of RFI in these
records is complicated by the lack of frequency resolution.  For BON
data we first carefully removed RFI in the time--frequency plane and
then used the radiometer equation, along with the known telescope noise
characteristics, to convert the rms noise fluctuation in the off-pulse
part of the profiles to a Jansky scale.  We determined this conversion
factor for the horizontal polarization and applied it to both channels.
We then added the corresponding profiles from both polarizations and, by
integrating the area under each profile, finally measured the pulsed flux
density.  On most days the flux densities estimated from the two systems
agree to within $\sim 25\%$, and hereafter we use only the BON values.

It is evident from Figure~\ref{fig:profs_nancay} that the pulse profiles
are not stable, particularly for the 2 months following late 2006 July
(MJD~53935 is July 19).  This presents obvious difficulties for obtaining
pulse times-of-arrival (TOAs) by following the usual prescription
of cross-correlating a ``standard profile'' with each day's profile.
Instead, we obtain TOAs from the maximum of a parabola fitted to one pulse
profile component that, we assume, regardless of variations, corresponds
to the same longitudinal fiducial point on the pulsar.  We choose for
this fiducial point the left-most of the (typically) two components that
make up the main cluster of emission.  This choice is motivated by the
facts that, within our large and densely sampled set of observations,
(1) the longitudinal separation between the corresponding peaks ($0.08P$)
remains essentially fixed even as the profiles vary enormously (e.g., see
the profiles for MJDs~53929 and 53977 in Fig.~\ref{fig:profs_nancay}, for
which the fiducial points are effectively the peaks of, respectively, the
highest and lowest clearly visible pulse components), and (2) the angular
separation between the left-most peak and the ``precursor'' visible
on some days 0.3 in pulse phase preceding it (which is generally much
fainter than the main components, but not always: see, e.g., the profiles
for MJDs~53925 and 53951) also remains constant within the uncertainties.
Also commending this method is the empirical consideration that it works
well, with only a handful of TOAs rejected because their pre-fit residuals
are particularly large, owing presumably to ``awkward'' pulse shapes.
The TOAs thus obtained for timing purposes (\S~\ref{sec:timing}) also
match (up to a constant offset of $\approx 25$\,ms) those obtained via
the standard cross-correlation method for data acquired before mid 2006
July (when profile variation was relatively small).

\subsubsection{Parkes}\label{sec:parkes}

The first detection of radio pulsations from \xte\ took place at the
ATNF Parkes telescope on 2006 March 17, with confirmation following on
April 25 \citep{crh+06}.  Since then we have observed the pulsar there at
frequencies ranging from 0.7 to 8.4\,GHz using a variety of spectrometers.
Unless otherwise noted, in this paper we use only 1.4\,GHz data, obtained
with the analog filterbanks as described by \citet{crh+06}.

The 1.4\,GHz Parkes data set is very sparse by comparison with \nancay's,
but is particularly useful here for two purposes: to extend the overall
timing span by more than 1 month, and to perform consistency checks on
the timing solution using two independent sets of TOAs.  Because the vast
majority of the Parkes data used here were obtained before late 2006 July
(when pulse shapes were relatively stable; see Fig.~\ref{fig:profs_nancay}
and \S~\ref{sec:nancay}), the standard cross-correlation method that we
used to obtain these TOAs works well enough.

\subsubsection{GBT}\label{sec:gbt}

\begin{figure}
\begin{center}
\includegraphics[angle=0,scale=0.46]{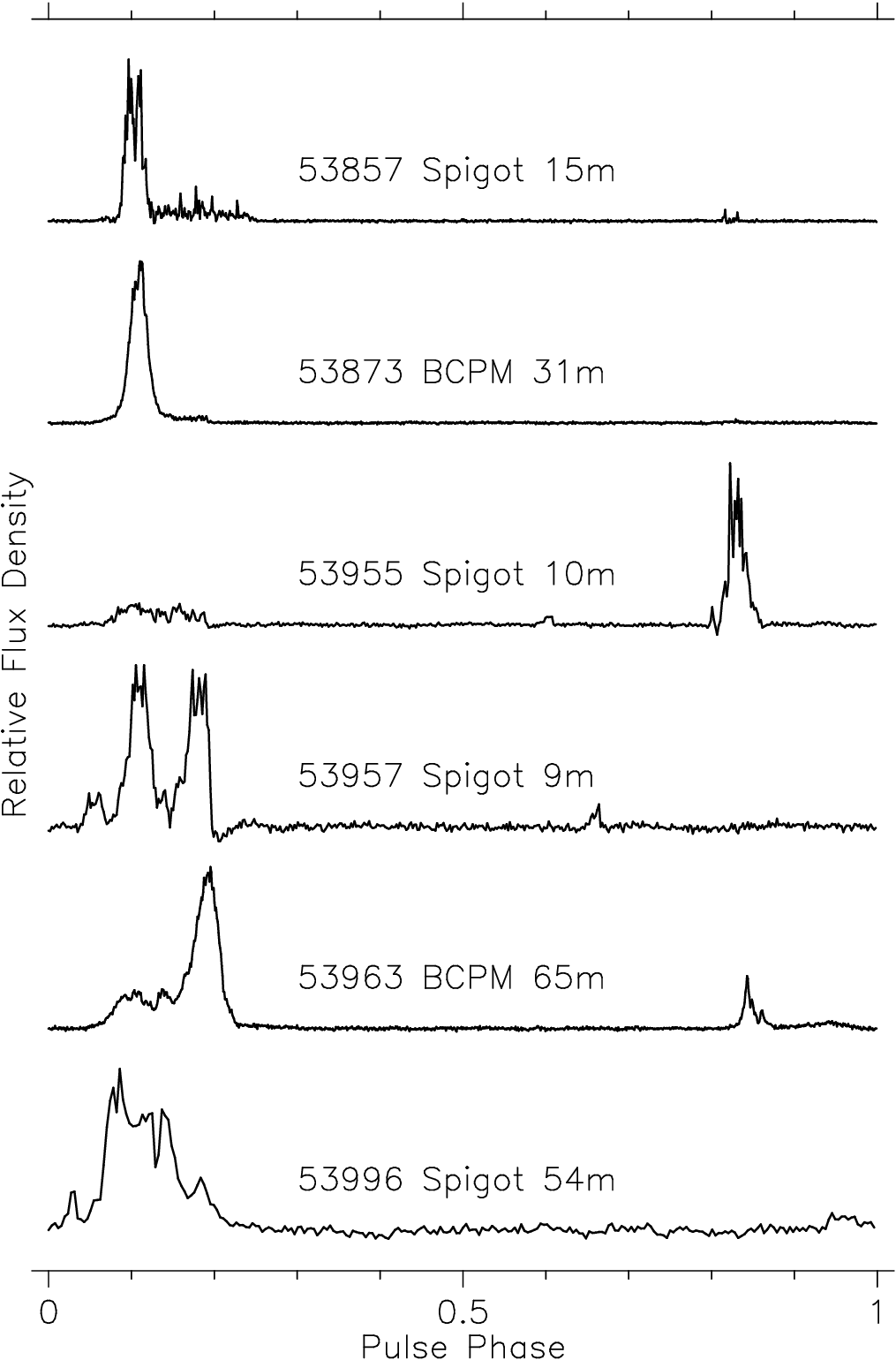}
\caption{\label{fig:profs_gbt} 
Changing pulse shapes of \xte\ from GBT observations.  All profiles
were aligned with the polynomial used for Fig.~\ref{fig:profs_nancay}.
The labels list the observation date (MJD), spectrometer used, and
effective integration time in minutes.  Spigot \citep{kel+05} data
were recorded across a 600\,MHz band centered at 1.95\,GHz, while a
134\,MHz band centered on 1.85\,GHz was used for BCPM \citep{bdz+97}.
Note the low-level emission near phases 0.8 on MJDs~53857, 0.6 on 53955,
0.65 on 53957, and 0.95 on 53963.  See also Fig.~\ref{fig:cxo}.  }
\end{center}
\end{figure}

We have observed \xte\ at the NRAO Green Bank Telescope (GBT) since
2006 May 2 at frequencies spanning 0.3--42\,GHz.  Very few of these
data are at 1.4\,GHz.  Starting in 2007 January, after the monitoring
rate decreased substantially at \nancay\ (\S~\ref{sec:nancay}), we
increased GBT monitoring at 1.9\,GHz, with data collected as described
by \citet{crh+06}.  The TOAs derived from these data, using the standard
cross-correlation method with pulse profiles that vary little, are used
to complement \nancay\ TOAs.  Also, using the large gain, bandwidth,
and frequency agility of GBT, we have obtained very high-quality pulse
profiles at a number of widely-separated frequencies that illustrate the
remarkable time-variation in pulse shapes displayed by this magnetar.
Some examples of this collection at a frequency of 1.9\,GHz are shown
in Figure~\ref{fig:profs_gbt}.

\subsubsection{VLA}\label{sec:vla}

We have used a sub-array of the NRAO Very Large Array (VLA) to measure
with high precision the flux density of \xte\ at 1.4\,GHz on 13 days
between 2006 February 28 and September 5.  During a typical observation
we accumulate 40\,min of on-source visibilities in continuum mode,
with a bandwidth of 100\,MHz.  Each observation begins with a $\approx
5$\,min scan of a flux calibrator, either 3C286 or 3C48.  The remainder
of the observation consists of three $\approx 2$\,min scans of the
phase calibrator 1811--209 interspersed with two $\approx 20$\,min
on-source scans.  The recorded visibilities are calibrated and imaged
using standard procedures in the Astronomical Image Processing System.
Due to variations in the array configuration, the number of antennas
used, and the amount of observing time allocated, the errors in our flux
density measurements range between 0.2\,mJy and 1.0\,mJy.

\subsubsection{{\chandra}}\label{sec:cxo}

\xte\ was observed with the {\em Chandra X-ray Observatory\/} on 2006
September 10--11, using the back-illuminated S3 chip of the ACIS-S CCD
detector in TIMED/VFAINT mode with a sub-array readout that provided a
time resolution of 0.441\,s, and deadtime of 9\%.  Data were acquired
continuously over an interval of 8.4\,hr starting at 19:40 UT with no
background contamination.  All photon arrival times were corrected to
the solar system barycenter (TDB) using the pulsar position given in
Table~\ref{tab:parms}, the JPL~DE200 ephemeris and the {\em axbary\/} task
in the Chandra Interactive Analysis of Observations.  Photons falling
within an aperture of radius $3''$ centered on \xte\ were extracted
from the standard processing event Level~2 data files and a total of
7014 counts were accumulated in the 0.5--3.0\,keV energy range, chosen
to maximize the source signal-to-noise ratio.

The barycentered photon arrival times were folded at epoch MJD~53989.0
(midnight TDB on September 11) into 10 bins using $P=5.540362$\,s
determined from a simultaneous radio observation at the GBT (see
\S~\ref{sec:phasing}).

\subsection{Timing}\label{sec:timing}

We have used the available TOAs
(\S\S~\ref{sec:nancay}--\ref{sec:gbt}), along with the TEMPO\footnote{See
http://www.atnf.csiro.au/research/pulsar/tempo/} software, to obtain
the timing solution that describes the rotational history of \xte\
during 2006 May--2007 January.  Because of the changing pulse shapes and
heterogeneity of the available data (\S~\ref{sec:obs}), particular care
must be taken with this process.

First we used \nancay\ TOAs to obtain a solution beginning in 2006 June.
Starting with a handful of TOAs and an initial solution fitting only
for rotational phase and frequency $\nu = 1/P$, we subsequently added
$\dot \nu$ as needed.  We then added one TOA at a time, paying particular
attention to its pre-fit residual and corresponding pulse shape, as well
as the post-fit residuals and fit parameters.  In this way we eliminated
seven TOAs, from a set of 144, that had large ($\ga100$\,ms) pre-fit
residuals; these could usually be ascribed to particularly unusual
pulse shapes.  After approximately 1 month it is no longer possible to
obtain featureless timing residuals without fitting for higher frequency
derivatives.  After this we added the Parkes TOAs, fitting also for
an offset between both TOA sets to account for an arbitrary alignment
between the respective fiducial points.  The magnitude of this offset
($21\pm15$\,ms; unless otherwise stated, the uncertainties used in
this paper all represent $1\,\sigma$ confidence levels) is as expected,
considering the different methods used to obtain the respective TOAs (see
\S\S~\ref{sec:nancay} and \ref{sec:parkes}).  Finally, when measuring
$\dot \nu$ in piece-wise fashion for 2007 January, we complemented the
\nancay\ TOAs with those from the GBT (\S~\ref{sec:gbt}).

In general it is difficult to estimate TOA uncertainties reliably for
our data, owing to changing pulse shapes and the principal method used
to obtain TOAs (see \S~\ref{sec:nancay}).  For the early Parkes data
these are typically in the 5--10\,ms range.  The uncertainty of early
\nancay\ TOAs is more often 10--20\,ms.  Following the general weakening
of the pulsar, the broadening of its profile components, and the huge
variability displayed by the pulse shapes (Fig.~\ref{fig:profs_nancay}),
we estimate that the TOA uncertainties are commonly around 25--50\,ms
but can reach values approaching 100\,ms on occasion.  As noted above
we have excluded from the timing fits a few TOAs with such large pre-fit
residuals (and estimated uncertainties).  Because of these difficulties,
we assigned a uniform weight to the TOAs and, as a result, we do not
obtain a $\chi^2$ figure of merit for the timing fits.  Nevertheless, the
TOA uncertainty estimates noted here are in keeping with the corresponding
fit rms residuals.

\begin{figure}
\begin{center}
\includegraphics[angle=270,scale=0.30]{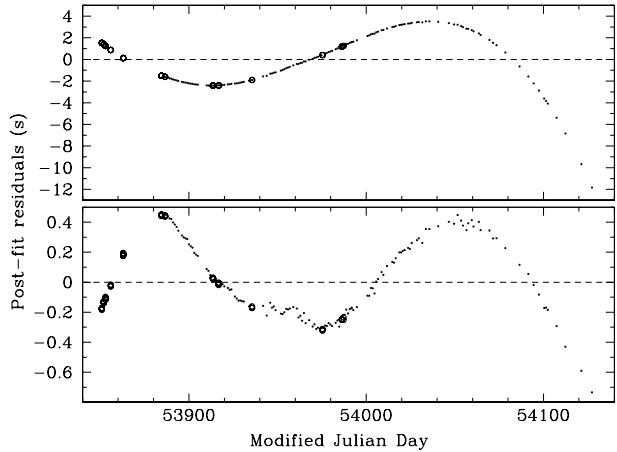}
\caption{\label{fig:res} 
Timing residuals for \xte.  Black dots correspond to \nancay\ TOAs, while
open circles represent Parkes TOAs (see \S~\ref{sec:obs}).  \textit{Top}:
Residuals for a model that contains only the rotation phase, frequency
$\nu$ and $\dot \nu$ (see Table~\ref{tab:parms}), showing a clear cubic
residual trend.  \textit{Bottom}: Residuals for a model with an additional
polynomial term, $\ddot \nu = 9.40\pm0.06 \times10^{-21}$\,s$^{-3}$.
Note the change in vertical scale by a factor of about 10, and the
remaining quartic trend.  }
\end{center}
\end{figure}

We obtained a phase-connected timing solution that encompasses
all data beginning with the confirmation TOAs from April 25 (we
comment below on use of the discovery TOAs from 2006 March 17).
The residuals from a quadratic fit to pulse phase are shown in the
top panel of Figure~\ref{fig:res} and the respective solution is given
in Table~\ref{tab:parms}.  An independent solution using only Parkes
TOAs results in similar parameters.  The cubic trend of large amplitude
visible in the residuals points to very significant unmodeled behavior,
expected of young pulsars in general (whose rotation is ``noisy'') and
of \xte\ in particular \citep{ims+04}.  While $\dot \nu$ clearly varies
with time, the bottom panel of Figure~\ref{fig:res} shows that it does
not do so smoothly.  A total of 11 frequency derivatives are required to
``whiten'' the residuals, which then have an rms of 18\,ms.

\begin{deluxetable}{p{3cm}l}
\tablewidth{0.70\linewidth}
\tablecolumns{2}
\tablecaption{\label{tab:parms} Timing parameters for \xte}
\tablehead{
\colhead{Parameter}  & \colhead{Value}}
\startdata
Right Ascension \dotfill                    & $18^{\rm h}09^{\rm m}51\fs087$  \\
Declination \dotfill                        & $-19\arcdeg43\arcmin51\farcs93$ \\
DM (cm$^{-3}$\,pc) \dotfill                 & 178.0                           \\
Epoch (MJD) \dotfill                        & 54000.0                         \\
$\nu$ (s$^{-1}$)~\tablenotemark{a} \dotfill & 0.180493891(6)                  \\
$\dot \nu$ ($10^{-13}$\,s$^{-2}$)~\tablenotemark{a} \dotfill & --2.53(1)      \\
MJD range \dotfill                          & 53850--54127                    
\enddata
\tablecomments{The celestial coordinates were held fixed at the values
obtained from VLBA observations \citep{hcb+07}, and the DM was held fixed
at the value obtained from simultaneous 0.7 and 2.9\,GHz observations
\citep{crh+06}. }
\tablenotetext{a}{These two parameters are sufficient to obtain a
phase-connected solution encompassing the MJD range, but do not fully
describe the rotation of the neutron star.  They are non-stationary and,
strictly, not predictive.  See Fig.~\ref{fig:res} and \S~\ref{sec:timing}
for more details. }
\end{deluxetable}

In order to extract from this record of unsteady rotation a quantitative
measure of the varying torque acting on the neutron star, we have measured
$\dot \nu$ as a function of time.  We have done this in a piece-wise
fashion, performing quadratic fits to phase for TOA spans long enough
that the nominal resulting fractional uncertainty in $\dot \nu$ was no
more than 3\%, but short enough that no trends were seen in the residuals.
In practice this implied individual segments of very nearly 30 days each.
So as to better sample the variation in $\dot \nu$, we did this with
\nancay\ data (supplemented in 2007 January by GBT TOAs) while stepping
through the TOAs in offsets of 15 days.  A fit to all the good \nancay\
TOAs using 11 frequency derivatives also yields $\dot \nu(t)$ (without
associated uncertainties) consistent with the discrete measurements.

Owing to the lack of \nancay\ observations and the sparser data from
Parkes and GBT, this was not straightforward before 2006 June 1.
We opted for performing timing fits using multi-frequency data from
both Parkes and GBT, thereby gaining some additional leverage in our
solutions.  At this time, the radio emission at most frequencies was
usually dominated by a single narrow component \citep[see][]{crh+06}, so
that TOA artifacts were relatively small.  In this manner, we obtained
a $\dot \nu$ measurement for late April--late May.  Finally, we also
used the 2006 March 17 Parkes discovery observations in a timing fit.
We appear to have maintained phase connection with the confirmation
observations 39 days later (the pre-fit residual between the discovery
TOAs and the solution computed with the first month of data following
confirmation is only $0.16P$).  However, we have established with \nancay\
data that cubic trends typically become significant for $\ga 30$ days,
and we have therefore obtained this earliest measurement of $\dot \nu$
also by computing the difference in barycentric frequencies between
2006 March 17 and April 25.  The two $\dot \nu$ values differ by only
$2.7\,\sigma$ of the (much smaller) uncertainty of the phase-connected
fit, and this agreement allows us to extend our record of $\dot \nu$
by 1 additional month.

\begin{figure}
\begin{center}
\includegraphics[angle=0,scale=0.40]{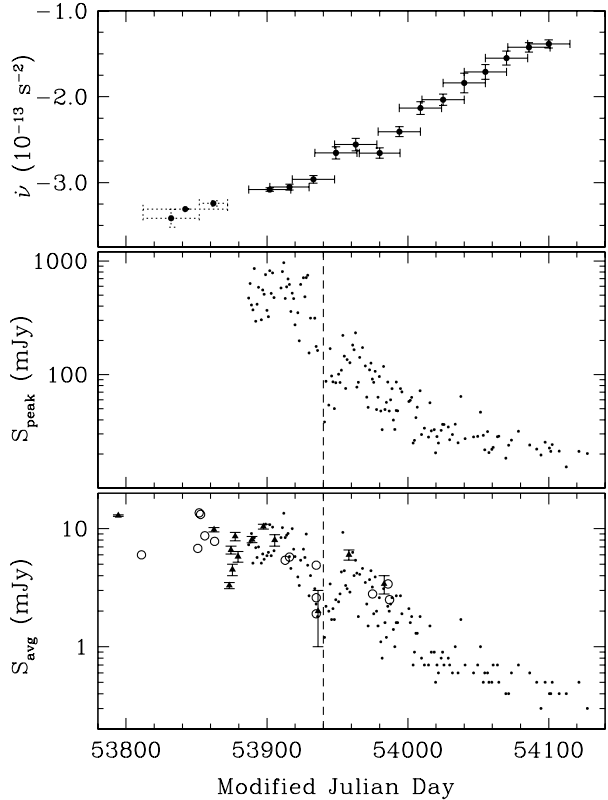}
\caption{\label{fig:fdot} 
Frequency derivative and flux densities at 1.4\,GHz for \xte.
\textit{Top}: Frequency derivative obtained from a variety of timing
fits each spanning about 30 days.  Solid error bars denote fits obtained
with \nancay\ data alone (excepting the last one, which also contains
some GBT data), while dotted error bars correspond to fits that contain
Parkes and in some cases also GBT data.  For fits containing \nancay\
data the error bars represent nominal $2\,\sigma$ confidence levels; for
all others, $4\,\sigma$.  In late 2006 July (indicated by the dashed line
at MJD~53940 in the lower panels), $\dot \nu$ increased at a large rate,
the flux densities reached a local minimum, and the pulse profiles changed
in character.  See \S~\ref{sec:timing} for details.  \textit{Middle}:
Daily peak flux density of the profile component used to obtain \nancay\
TOAs (see \S~\ref{sec:nancay} and Fig.~\ref{fig:profs_nancay}).
\textit{Bottom}: Daily period-averaged flux density from \nancay\
(\textit{black dots}) and Parkes (\textit{open circles}), and continuum
VLA flux density (\textit{triangles with error bars}). }
\end{center}
\end{figure}

The run of $\dot \nu$ measured over 9 months is shown in the top panel
of Figure~\ref{fig:fdot}.  The overall trend is one of increasing $\dot
\nu$ (as could be inferred from the top panel of Fig.~\ref{fig:res})
but the way in which this occurs is far from steady.  In the 3
months prior to mid 2006 July, $\dot \nu$ varied in a relatively
steady and gradual fashion, from $-3.3\times10^{-13}$\,s$^{-2}$ to
$-3.0\times10^{-13}$\,s$^{-2}$.  Then, in late July, it changed to
$-2.7\times10^{-13}$\,s$^{-2}$ over a span of 15 days --- the implied
change in torque is $-2\times10^{32}$\,dyn\,cm (using a stellar moment
of inertia of $10^{45}$\,g\,cm$^2$), which is larger than the torque
powering three-quarters of all known ordinary pulsars!  Regardless of
the detailed mechanisms that produce radio pulses in \xte, it should
perhaps not be surprising if other observational properties of the
magnetar should have changed around this time.

In fact, as the middle panel of Figure~\ref{fig:fdot} shows, the peak
flux density of \xte\ has dramatically decreased since about that
time.  Interestingly, while the period-averaged flux density has also
decreased compared to its average value before mid July (bottom panel
of Fig.~\ref{fig:fdot}), until about 2006 October it did so by a smaller
factor and continued to fluctuate greatly from day to day (when average
flux densities are available from at least two of \nancay, Parkes or
VLA within a day of each other, they are consistent within expectations
given the inherent variations).  These two observations can be understood
by inspection of the profiles shown in Figure~\ref{fig:profs_nancay}:
from late July to mid September ($\mbox{MJD} \sim 53994$), the daily
profiles of \xte\ tended to be composed of two (or more) significant
peaks, each much broader than the one peak generally prominent before
mid July (with typical full-width at half-maximum $\approx0.04P$, versus
about half that value beforehand), and with greater pulse-shape variance
than before.  This may also explain why the \nancay\ timing residuals
for each 1-month fit after mid 2006 July are about 24\,ms rms, twice as
large as the typical corresponding value before then.

Since 2006 October, the \nancay\ profiles appear to have varied less, and
to be composed mainly of one broad peak (Fig.~\ref{fig:profs_nancay}),
although on some days the trailing peak is still recognizable (as it is
more often in higher quality GBT data).  After another relatively large
increase over the month of September, $\dot \nu$ has continued a steady
increase at approximately the average rate for the 9-month span of our
observations, $7.5\times10^{-16}$\,s$^{-2}$ per day.

\subsubsection{X-ray and radio pulse alignment}\label{sec:phasing}

\begin{figure}
\begin{center}
\includegraphics[angle=0,scale=0.46]{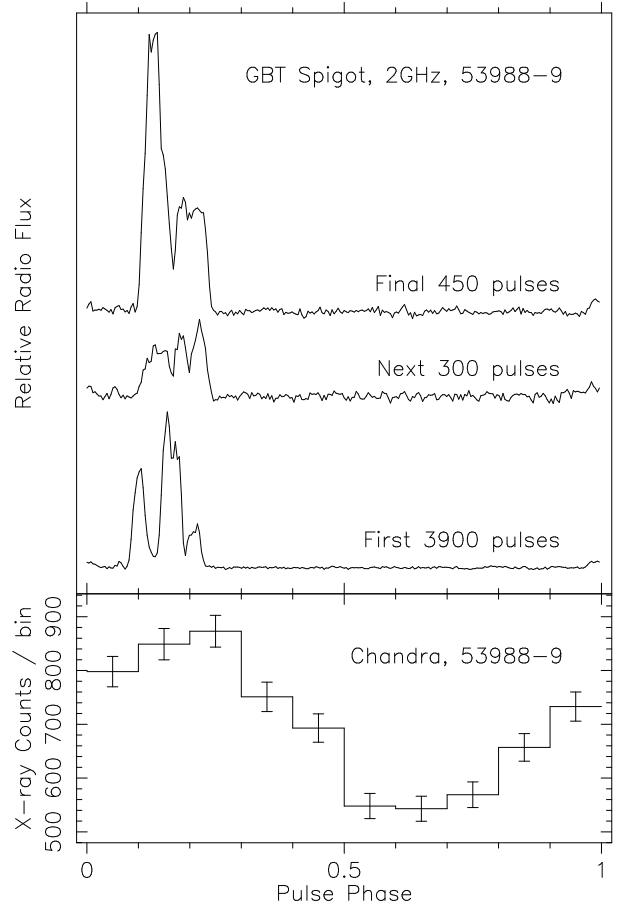}
\caption{\label{fig:cxo} 
\textit{Bottom}: X-ray pulse profile (0.5--3.0\,keV) of \xte\ from
an observation with \chandra\ on 2006 September 10--11, folded with
barycentric $P=5.540362$\,s as determined from a simultaneous radio
observation (see \S~\ref{sec:cxo}).  Phase zero on this plot corresponds
to MJD~53989.0 (TDB).  \textit{Top}: Radio profiles from a simultaneous
GBT observation, starting on September 10, 20:50 UT.  After 6\,hr
(\textit{bottom trace}), the pulse profile abruptly changed to a different
configuration for 30\,min (\textit{middle trace}), after which it changed
yet again (\textit{top trace}).  The relative integrated flux densities
of the three profiles are 1.4, 1.0, 2.9, from bottom to top.  The middle
peak of the bottom radio profile arrives at phase 0.17 on the plot, and
all four profiles are absolutely aligned (see \S~\ref{sec:phasing}).
Note also the alignment between the left-most peak of the top profile
and the first notch in the bottom profile.  }
\end{center}
\end{figure}

On 2006 September 10--11 we observed \xte\ for 7.2\,hr with the
GBT at 1.9\,GHz, starting 1\,hr after the beginning of the \chandra\
observation (\S~\ref{sec:cxo}).  We have used the contemporaneous TOAs
thus obtained to measure the phase offset between the X-ray and radio
pulses.  The folded X-ray profile is shown in Figure~\ref{fig:cxo},
with the first phase bin chosen to begin at midnight on September 11
(TDB).  After translating the TOAs for the first radio profile (see
Fig.~\ref{fig:cxo}) converted to infinite frequency to the solar system
barycenter, a TEMPO fit yields a phase offset for the fiducial point of
the radio profile (here its middle peak) of $0.167\pm0.006$ in the Figure.
The uncertainty includes a component due to the fit and a slightly smaller
contribution from the uncertainty in the dispersion measure (DM) of \xte.

Within the larger uncertainty imposed by the \chandra\ resolution
($0.08P$; see \S~\ref{sec:cxo}) and the relatively small number of pulsed
X-ray counts, the main component of the radio profile on this day arrives
at the same time as the peak of the X-ray pulse (Fig.~\ref{fig:cxo}).

Interestingly, the radio profile changed in both shape and flux
during the \chandra\ observation --- twice, within a span of 30\,min
(Fig.~\ref{fig:cxo}).  This does not modify the conclusion in the above
paragraph, but helps answer a question concerning the radio flux and
profile variations observed in \xte: they can occur suddenly (observed at
a resolution of $\sim 10$\,s).  We have ``caught in the act'' at least
nine such changes at \nancay, GBT and Parkes, all since mid 2006 July.
As this corresponds to some 150 hours of observing time, we can estimate
that such transitions occur on average every $\sim 15$\,hr, at the
present epoch.  No X-ray bursts or significant changes in the X-ray
flux or pulse shape were seen at the times of the radio transitions
indicated in Figure~\ref{fig:cxo}.  The \chandra\ count rates during
the three radio pulse states --- $0.235 \pm 0.003$, $0.213 \pm 0.011$,
and $0.218 \pm 0.007$\,s$^{-1}$, respectively --- are uncorrelated with
the large changes in radio flux density and pulse shape.

\section{Discussion}\label{sec:disc}

All young radio pulsars experience rotational instabilities, observed
as a continuous quasi-stochastic wandering of the pulse phase (``timing
noise'') or as discontinuous spin-up events (``glitches'').  These are
thought ultimately to be driven by the unsteady transfer of angular
momentum from the interior superfluid to the crust of the neutron star.
\xte\ displays large amounts of timing noise-like behavior; despite
very large changes observed in $\dot \nu$ on short time scales, all
observed rotational changes have been continuous to within the available
resolution and we have not observed any clear glitch-like behavior.
\citet{antt94} quantified timing noise as the time-magnitude of its
cumulative contribution over a time interval $t$ to the cubic term
in a Taylor series expansion of rotational phase, i.e., $\ddot \nu
t^3/(6 \nu)$.  For \xte\ this amounts to 120\,s over 277 days, a huge
amount that far surpasses anything observed in radio pulsars and is also
greater than an extrapolation based on the notion that the magnitude
of timing noise is proportional to $\dot P$ \citep{antt94}.  However,
this level of timing noise is not unprecedented for magnetars.

Six AXPs, including \xte, and two soft-gamma repeaters (SGRs)
--- i.e., the majority of all known magnetars --- have been
observed with phase connection maintained for at least a few months
\cite[see][]{kgc+01,ggk+02,gk02,wkg+02,kg03,gk04}.  The SGRs~1806--20 and
1900+14 usually have the largest amounts of timing noise, roughly one
order of magnitude above the level we observe in \xte, with $\dot \nu$
varying by factors of up to 4 \citep{wkg+02}.  The rotational stability of
AXPs is generally greater \citep[although at least two have been observed
to glitch; see][]{kgw+03,kg03}, and four are substantially quieter than
\xte.  The remaining one, 1E~1048.1--5937, was relatively quiet during
2005--2006 \citep{kas06}, but previously displayed far greater rotational
instability than observed in \xte\ \citep{gk04}.  Therefore, while noisier
than most AXPs, \xte\ may not be particularly remarkable in this respect.
What is unusual is that we can now track its rotation via nearly daily
observations, in greater detail than is possible for other AXPs.

Following its discovery with the {\em Rossi X-ray Timing
Explorer}, a phase-connected solution was obtained for \xte\
spanning 2003 January--September, during which $\dot \nu$ varied
between $-6.7\times 10^{-13}$\,s$^{-2}$, averaged over 8 months, and
$-3.8\times 10^{-13}$\,s$^{-2}$ over the last 2 months \citep{ims+04}.
The very large initial magnitude may have reflected in part transients
associated with a putative glitch at the time of the X-ray outburst in
early 2003.  In the succeeding 2.5 years, until the radio detection,
phase connection was no longer possible owing to the sparse sampling,
but individual period measurements 6 months apart established, e.g.,
that the average $\dot \nu$ was $-1.7\times 10^{-13}$\,s$^{-2}$ between
late 2003 and early 2004 \citep[see][]{gh05}.  The value of $\dot \nu =
-1.4\times 10^{-13}$\,s$^{-2}$ in early 2007 is the largest ever measured
for this source, for which the historical range can be summarized as
$\dot \nu = (- 2.6 \pm 1.2) \times 10^{-13}$\,s$^{-2}$ spanning late
2003--early 2007.  Importantly, although $\dot \nu$ has increased almost
monotonically during our radio observations (Fig.~\ref{fig:fdot}),
it cannot have been increasing during the entirety of the past 3 years
\citep[see also][]{gh06}.  Periods of decreasing torque such as we have
observed recently must have been interspersed with at least one epoch of
substantially increasing torque.  If the torque for \xte\ were to continue
decreasing at the rate in 2007 January, by mid 2007 the pulsar would
stop spinning down altogether; presumably this rate will soon decrease.

In ordinary pulsars, the observed rotational instabilities are mainly
presumed to be caused by angular momentum transfer processes internal
to the neutron star.  In the magnetar model, the external energy
stored in the magnetosphere, partly released through reconfiguration
and decay of the magnetic field and currents, may also lead to erratic
variations in the torque acting upon the star, for example via a flux
of Alfv\'{e}n waves and particles resulting from magnetically-induced
(sudden large-scale or persistent small-scale) crustal seismic activity
\citep[e.g.,][]{tdw+00}.  It is unclear whether any such particular models
\citep[see also][]{dun01} can explain the magnitude and time scale of
the variations in torque currently observed in \xte, 4 years after the
X-ray outburst.  In any case, substantial variations in magnetospheric
plasma densities and/or currents would likely have implications for
other observable properties of the pulsar.

An interesting comparison is with PSR~B1931+24, an ordinary
radio pulsar during intervals of 5--10 days, with $P=0.8$\,s and
$B\approx2\times10^{12}$\,G, which abruptly shuts off for 25--35 days in
a pattern that repeats quasi-periodically \citep{klo+06}.  During these
turned-off periods, the torque is only two-thirds of its turned-on
value.  \citet{klo+06} conclude from this extraordinary behavior that
the occasional presence of plasma leads to radio emission and its flow
provides the extra braking torque.  The plasma density calculated from
the torque difference is 100\,esu\,cm$^{-3}$, in agreement with the
co-rotation value \citep{gj69}.  In \xte\ the radio emission has not
shut off, but it did diminish and change markedly in character somewhat
abruptly in late 2006 July, apparently coincident with a huge reduction
in torque ($\dot \nu$ changed from $-2.96\times10^{-13}$\,s$^{-2}$
on MJD~53933 to $-2.65\times10^{-13}$\,s$^{-2}$ on MJD~53949; see
\S~\ref{sec:timing} and Fig.~\ref{fig:fdot}).  A calculation such as that
by \citet{klo+06} would suggest for \xte\ an ``extra'' plasma density of
about 300\,esu\,cm$^{-3}$ prior to the large torque decrease, compared
to the source's Goldreich-Julian density of 1350\,esu\,cm$^{-3}$.
Implicit in this calculation is the notion that the radio emission
originates from open field lines, which has not been proven for \xte.
For a magnetar, there is additional plasma due to non-axisymmetric
large-scale magnetospheric currents \citep{tlk02}.  In any case,
the pulse profile and flux variations observed in \xte\ occur on time
scales that are apparently too short to be explained by changes in the
closed field lines \citep[see, e.g.,][]{bt06}.  Also, interpretation of
polarimetric data for \xte\ appears consistent with emission from open
field lines \citep{crj+07}.

If indeed the measured changes in torque are caused by variations in the
magnetospheric plasma density in locations relevant for the production
of coherent radio emission, then it should be no surprise that the
pulse profiles of \xte\ change so remarkably: even the ``small'' torque
changes reflected in Figure~\ref{fig:fdot} are enormous by the standards
of ordinary pulsars.  And certainly, the profile variations in \xte\
(Figs.~\ref{fig:profs_nancay}, \ref{fig:profs_gbt} and \ref{fig:cxo})
are observationally distinct from the ``mode changing'' of some ordinary
pulsars, where the average pulse profile suddenly changes between two
of a small set of different configurations \citep[e.g.,][]{bmsh82};
\xte\ displays a much greater variety.  However, the root cause of
mode changing is not well understood and, as we have established that
\xte\ profiles can change suddenly (Fig.~\ref{fig:cxo}), some link
may exist between these two phenomena.  Other behavior also appears
extraordinary, such as the broader pulse components typically observed
after MJD~53940 even as the phase separation between components remains
fixed (Fig.~\ref{fig:profs_nancay} and \S~\ref{sec:timing}) and the
polarimetric properties remain largely unchanged \citep{crj+07}.

Compared to such extreme radio variability, the X-ray spectrum and pulse
profiles of \xte\ change slowly as the flux decays \citep[e.g.,][]{gh06},
on a much longer time scale ($\sim 1$\,yr) than the fluctuations in
torque reported here.  The alignment of the peaks of the radio and X-ray
pulses suggests that the footpoints of the active magnetic field lines on
which radio emission is generated are also the locations of concentrated
crustal heating that is responsible for the enhanced X-ray emission,
at least at the higher energies that apparently come from a relatively
small area \citep[see][]{gh06}.  However, even if the energetic particle
bombardment that heats the surface hot spot fluctuates on time scales of
less than 1 day, the absence of correlated X-ray variability on similar
time scales indicates that most of the X-ray luminosity originated in
deeper crustal heating at the time of the X-ray turn on, or from more
gradual decay of the magnetic field \citep{ec89a}.

Among persistent AXPs it is not altogether clear what is the relationship
between observed radiative properties (such as pulse shapes and fluxes)
and rotational evolution (such as variations in torque).  Pulse profiles
can change following glitches \citep[e.g.,][]{kgw+03}, but otherwise
appear to be fairly stable \citep{gk02}.  Quite apart from the phenomenon
of X-ray bursts \citep[for bursts in \xte, see][]{wkg+05}, some AXPs have
shown substantial variability in X-ray flux \citep[e.g.,][]{gk04}.
Nevertheless, with the possible exception of 1E~1048.1--5937
\citep{gk04}, there has been no reported correlation between X-ray flux
and spin-down rate.  For \xte, on the other hand, there is an observed
correlation between radio flux density and torque over a period of 9
months (Fig.~\ref{fig:fdot}), although we know neither whether this
correlation is causal nor whether it holds over longer periods of time.

With observations of \xte, we now find ourselves in the curious position
of relying upon energetically insignificant radio pulsations exhibiting
a remarkably diverse phenomenology to illuminate rather more energetic
events on the magnetar.  Attempting to understand key aspects of the
radio observations may lead to a deeper understanding of both magnetars
and radio pulsar emission.

\acknowledgments

We are grateful to John Sarkissian for help with Parkes observations,
Eric Gerard for useful discussions concerning \nancay\ flux calibration,
and David Nice for wisdom on time systems as used in TEMPO.  The \nancay\
radio telescope is part of the Paris Observatory, associated with the
Centre National de la Recherche Scientifique (CNRS), and partially
supported by the Region Centre in France.  The National Radio Astronomy
Observatory is a facility of the National Science Foundation, operated
under cooperative agreement by Associated Universities, Inc.  The Parkes
Observatory is part of the Australia Telescope, which is funded by the
Commonwealth of Australia for operation as a National Facility managed
by CSIRO.  FC thanks the NSF for support through grant AST-05-07376.
EVG acknowledges support for this work provided by the National
Aeronautics and Space Administration through Chandra Award Number
GO6-7044X issued by the Chandra X-ray Observatory Center, which is
operated by the Smithsonian Astrophysical Observatory for and on behalf
of NASA under contract NAS8-03060; and by NASA ADP grant NNG05GC43G.

\end{document}